\documentclass{article}
\usepackage{arxiv}
\usepackage[utf8]{inputenc} 
\usepackage[T1]{fontenc}    
\usepackage{hyperref}       
\usepackage{url}            
\usepackage{booktabs}       
\usepackage{amsfonts}       
\usepackage{nicefrac}       
\usepackage{microtype}      
\usepackage{lipsum}
\usepackage{graphicx}
\usepackage{booktabs}       
\usepackage{amsfonts}       
\usepackage{nicefrac}       
\usepackage{microtype}      
\usepackage{lipsum}
\usepackage{graphicx}
\usepackage{amsmath,amssymb}
\usepackage{dcolumn}
\usepackage{bm}
\usepackage{braket}
\usepackage[dvipsnames]{xcolor}
\usepackage{xcolor,colortbl}
\usepackage{makecell,booktabs}
\usepackage{graphicx}
\usepackage{dcolumn}
\usepackage{listings}
\usepackage{bm}
\usepackage{braket}
\usepackage[dvipsnames]{xcolor}
\usepackage{xcolor,colortbl}
\usepackage{braket}
\usepackage{amsmath,amssymb,amsfonts}
\usepackage{algorithmic}
\usepackage{textcomp}
\usepackage{url}
\usepackage{bm}
\usepackage{times}
\usepackage{epsfig}
\usepackage{graphicx}
\usepackage{amsmath}
\usepackage{multirow}
\usepackage{amssymb}
\usepackage{subfig}
\usepackage{graphicx}
\usepackage{float}
\usepackage{booktabs}
\usepackage{textcomp}
\usepackage{adjustbox}\usepackage{mathtools}
\usepackage{array}
\usepackage{longtable}
\usepackage{makecell,booktabs}
\usepackage{multirow}
\usepackage{stackengine}
\usepackage{algorithm}
\usepackage{graphicx}
\usepackage{listings}
\usepackage{color}
\usepackage{color,array}
\usepackage{cite}
\usepackage{amsmath,amssymb,amsfonts}
\usepackage{algorithmic}
\usepackage{textcomp}
\usepackage{url}
\usepackage{bm}
\usepackage{times}
\usepackage{epsfig}
\usepackage{graphicx}
\usepackage{amsmath}
\usepackage{multirow}
\usepackage{amssymb}
\usepackage{subfig}
\usepackage{graphicx}
\usepackage{float}
\usepackage{booktabs}
\usepackage{textcomp}
\usepackage{adjustbox}\usepackage{mathtools}
\usepackage{array}
\usepackage{makecell,booktabs}
\usepackage{multirow}
\usepackage{stackengine}
\usepackage{algorithm}
\usepackage{graphicx}
\usepackage[dvipsnames]{xcolor}
\usepackage{xcolor,colortbl}

\graphicspath{ {./images/} }

\usepackage[title]{appendix}

\definecolor{amber}{rgb}{1.0, 0.75, 0.0}
\definecolor{orange}{rgb}{1.0, 0.49, 0.0}
\definecolor{codegreen}{rgb}{0,0.6,0}
\definecolor{codegray}{rgb}{0.5,0.5,0.5}
\definecolor{codepurple}{rgb}{0.58,0,0.82}
\definecolor{backcolour}{rgb}{0.95,0.95,0.92}

\lstdefinestyle{mystyle}{
    backgroundcolor=\color{backcolour},   
    commentstyle=\color{codegreen},
    keywordstyle=\color{magenta},
    numberstyle=\tiny\color{codegray},
    stringstyle=\color{codepurple},
    basicstyle=\ttfamily\footnotesize,
    breakatwhitespace=false,         
    breaklines=true,                 
    captionpos=b,                    
    keepspaces=true,                 
    numbers=left,                    
    numbersep=5pt,                  
    showspaces=false,                
    showstringspaces=false,
    showtabs=false,                  
    tabsize=2
}

\title{Treatment-Response Analysis of Tumor as A Quantum Particle}

\author{
    Nam Nguyen, M.A \\
    Department of Statistics\\
    University of South Florida\\
    Tampa, FL 33620 \\
    \texttt{namphuongnguyen510@gmail.com} \\
}

\begin{document}
\maketitle

\begin{abstract}
In this article, I present a novel and computational-efficient approach for treatment-response modeling of tumor progression-free survival (PFS) probability using the physical phenomenon of a quantum particle walking on a one-dimensional lattice with a proximate trap. The implementation is made available at: \url{https://github.com/namnguyen0510/Tumor-As-Quantum-Particle/tree/main}.
\end{abstract}

\section{Introduction}
\color{black}
Quantum computing is a new technology that promises a new way to accelerate or perhaps revise our look at the current Machine Learning (ML) models. It is noted on the current proposal for Geometric Deep Learning that any ML model is a group action on set\cite{bronstein2017geometric,bronstein2021geometric}; here, the set includes input signal, and the group of action can be considered as \textbf{functor}\cite{gratzer2008universal}. Current Euclidean embedding is representation in $GL_{n\times m}(\mathbb{R})$, while quantum machine learning models learn the embeddings on $GL_{n\times m}(\mathbb{C})$. Our previous works have addressed several applications\cite{nguyen2022quantum,nguyen2022bayesian} of quantum neural networks, which show agreement with the early literature of the field on the potential of using quantum computing to unlock the full potential of Artificial Intelligence.

\subsection{Quantum Neural Networks vs. Classical Neural Networks}
A classical neural network is given as a parameterized function
$$\hat{y} = f_{\bm{\theta}}(X, y);$$ where $X$ is the training input, $y$ is the label and $\hat{y}$ is the predicted values. The transformation of data in the classical model is presented as 
\begin{equation}
	\begin{split}
	f: GL(\mathbb{R}^p) \rightarrow GL(\mathbb{R}^q)\\
	X_{p=n \times m} \rightarrow X'_{q = n' \times m'}
	\end{split}
\end{equation}

Quantum neural networks instead using transformation on the Hilbert vector space $\mathcal{H}$ using these following rotation in Ox, Oy, Oz axis parameterized by $\theta_{ij}$ given by $$R_{\sigma_x}(\bm{\theta}) =  \begin{bmatrix}
\cos (\bm{\theta}/2) & -i \sin(\bm{\theta}/2)\\
-i \sin(\bm{\theta}/2) & \cos (\bm{\theta}/2)
\end{bmatrix},$$ $$R_{\sigma_y}(\bm{\theta}) =  \begin{bmatrix}
\cos (\bm{\theta}/2) & -\sin(\bm{\theta}/2)\\
\sin(\bm{\theta}/2) & \cos (\bm{\theta}/2)
\end{bmatrix},$$ and $$R_{\sigma_z}(\bm{\theta}) = \begin{bmatrix}
e^{-i\frac{\bm{\theta}}{2}} & 0\\
0 & e^{i\frac{\bm{\theta}}{2}}
\end{bmatrix}.$$
We give some comparisons between QNNs to some model architecture:
\begin{enumerate}
    \item QNNs can be considered as a capsule-neural network\cite{sabour2017dynamic, hinton2018matrix} if we consider the representation before the measurement of the quantum electronic wavefunction. The quantum neuron ("quron") encodes data by a complex-valued functor.
    \item Transformation inside quantum neural network has the geometric information of functor used, represented via Pauli-based (3D) transformations.
    \item QNNs can be viewed as physical-based machine learning because their representations are presented as electronic wavefunctions.
\end{enumerate}

\subsection{Contribution}
In this work, we aim to achieve the following goals:
\begin{enumerate}
    \item Translation of the theoretical model in \textbf{Equation}~\ref{equa:survival} for learning the non-linear dynamics of tumor evolution, discussed in \textbf{Section}~\ref{section:non_linear_tumor}.
    \item A loss module will be introduced in \textbf{Section}~\ref{sec:loss_module} for efficient training of the quantum model in the context of PFS prediction.
    \item We propose three ways to explain the model prediction. Of note, explainable AI is a hot topic in the current ML literature, as we also attempt to improve the model explainability in this work. Our model prediction delivers three main analyses:
    \begin{enumerate}
        \item Global observations of the entire cohort, including (1) the progression-free probability and (2) response score.
        \item Sub-class specific prediction - T.A.R.G.E.T plots, which quantizes prediction surfaces into different patient classes.
    \end{enumerate}
\end{enumerate}
I organize this article as follows: 
\begin{enumerate}
    \item Section~\ref{sec:_preliminary} introduces the preliminary.
    \item Section~\ref{sec:cls_quantum} gives the translation of the quantum models to model PFS of tumor progression.
    \item Section~\ref{sec:result} reports the numerical result, demonstrated on TCGA database.
    \item Section~\ref{sec:discussion} gives discussions and conclusion of this work.
\end{enumerate}
\color{black}

\color{black}

\section{Preliminary}\label{sec:_preliminary}
\subsection{Classical random walk vs Quantum random walk}
Classical random walk is a stochastic process in which a particle or walker moves randomly in a one-dimensional space, either to the left or right, with equal probability. At each time step, the particle moves one step in either direction with a fixed probability. The position of the particle after a certain number of steps is determined by the random outcomes of each step. The classical random walk is described by classical probability theory, and the walker's motion is deterministic and entirely predictable.

It is observed that a classical random walk follows a normal distribution (\textbf{Figure}~\ref{fig:cls_quan}(C)), centralized around the initial position. In contrast, quantum random walks have a diffused distribution for particle position as the tumor is in the superposition, meaning it is simultaneously progressing and not-progress at the same time before measurement. Of note, quantum information is collapsed into classical information after the measurement.

\subsection{Limitation of Kaplan–Meier Survival Analysis}\label{sec:limitation}
The Kaplan-Meier (KM) estimator\cite{kaplan1958nonparametric} is a non-parametric statistic used to estimate the survival function of a population or a group of individuals, one of the most dominant models in medical research. The model enables the estimation of the survival probability of patients, study participants, or systems over time. The KM estimator is given by 
\begin{equation}
    \hat{S}(t) = \prod_{t_i \le t} \frac{n_i - d_i}{n_i},
\end{equation}
where $\hat{S}(t)$ is the Kaplan-Meier estimate of the survival probability at time t, $t_i$ is the time of the $i^{th}$ event, $n_i$ is the number of individuals at risk just before the $i^{th}$ event, and $d_i$ is the number of individuals who experience the $i^{th}$ event.

\begin{figure}[t]
    \centering
    \includegraphics[width = 0.33\textwidth]{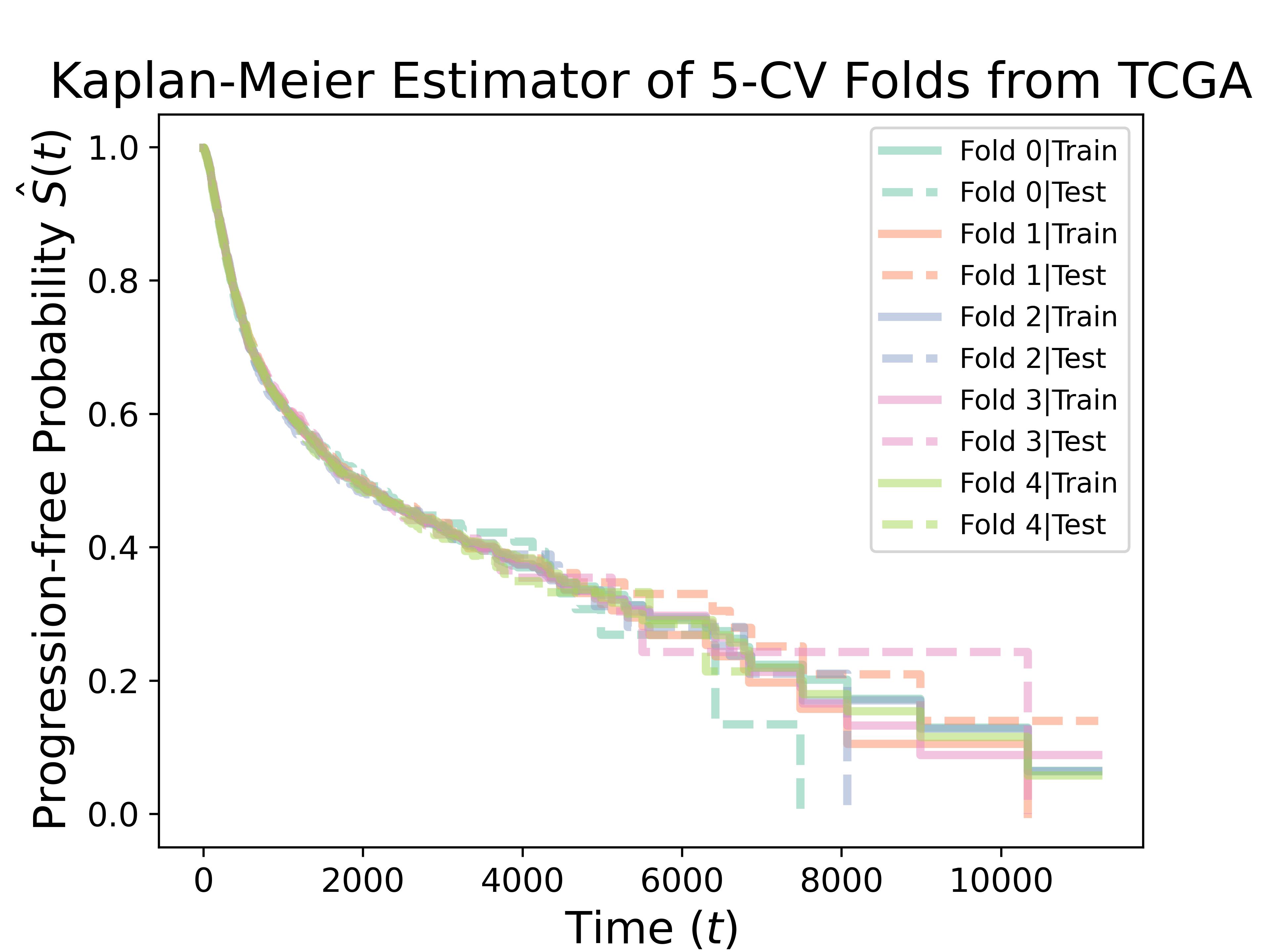}
    \includegraphics[width = 0.33\textwidth]{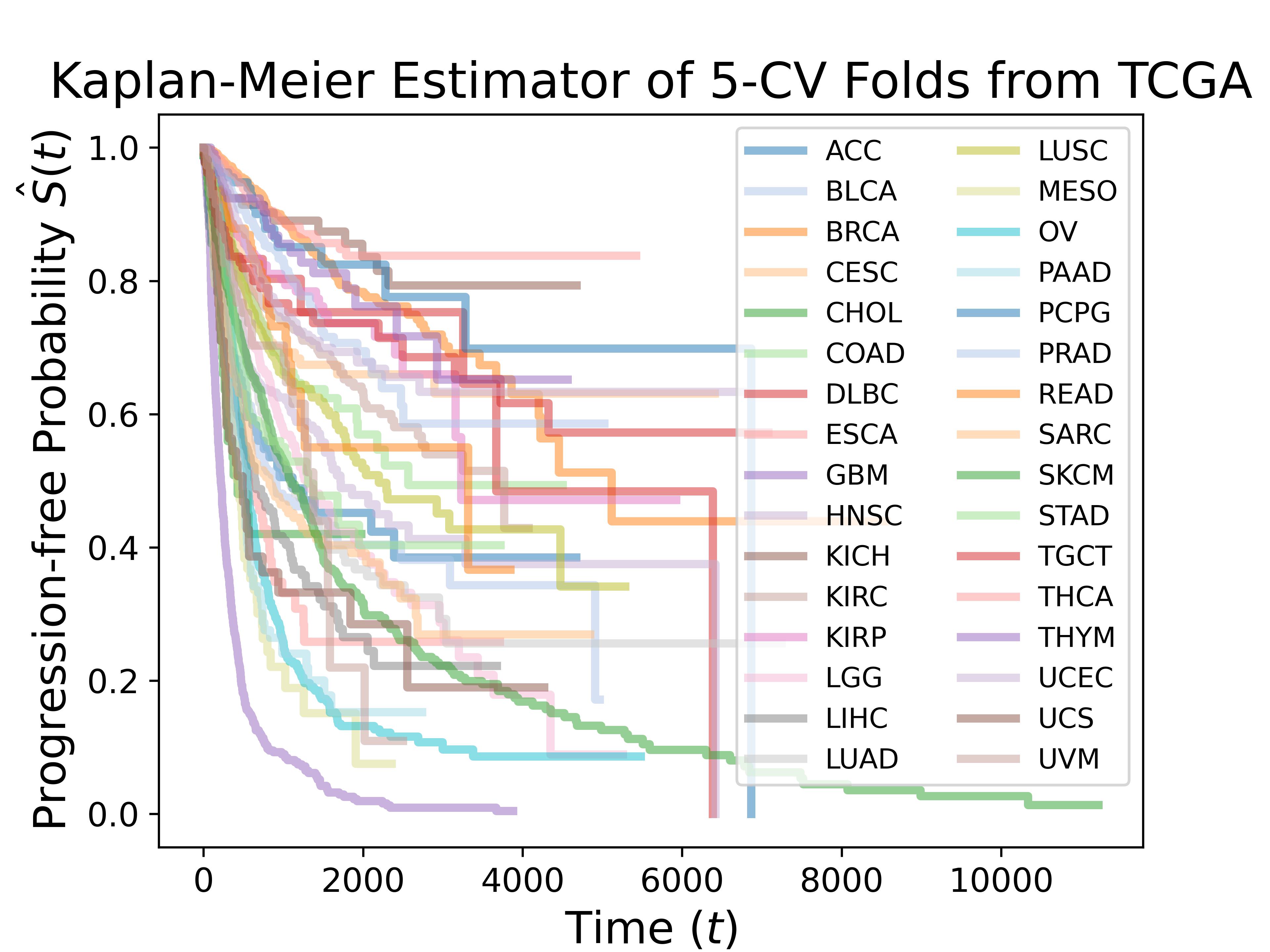}
    \includegraphics[width = 0.33\textwidth]{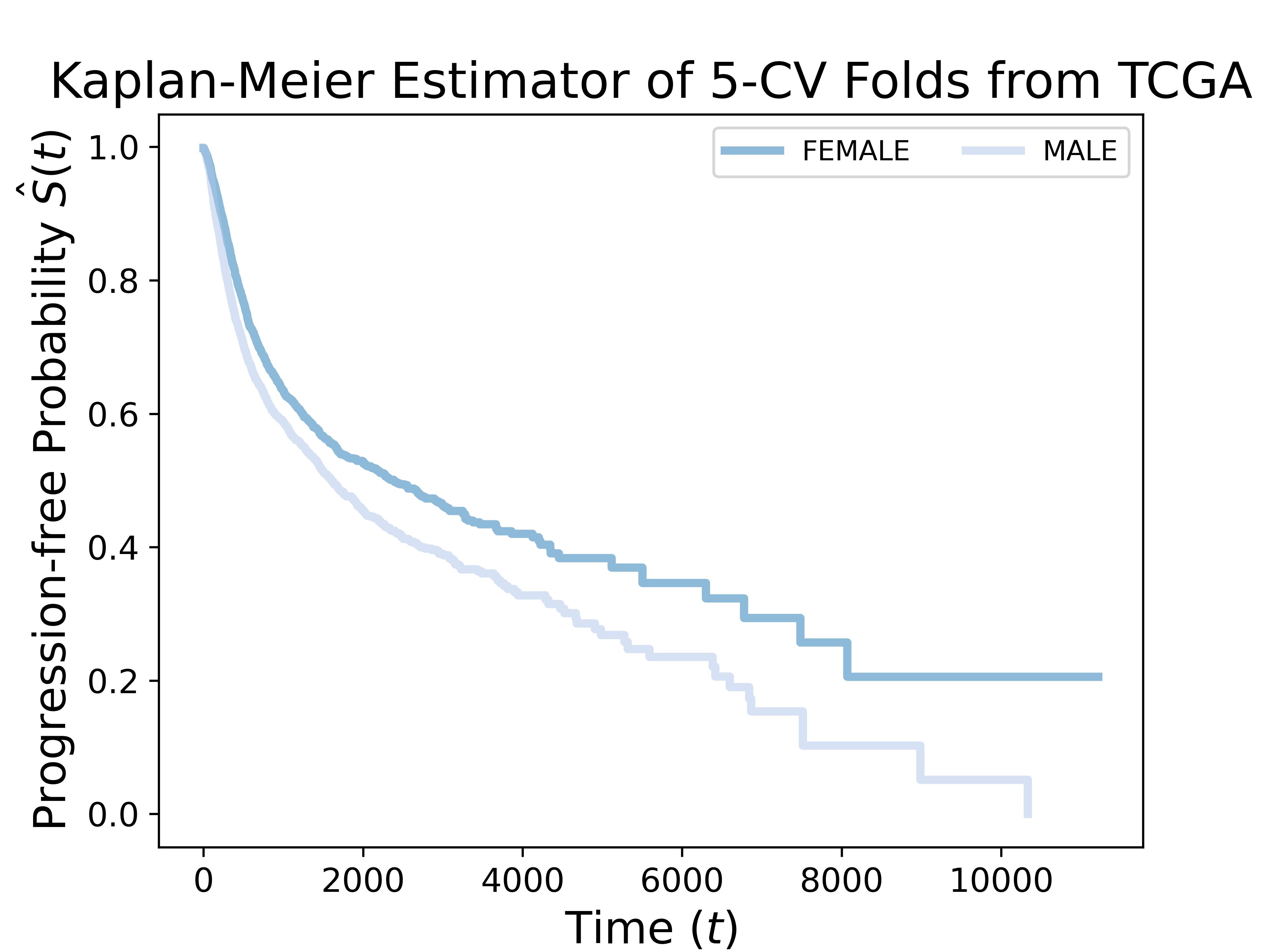}
    \caption{\textbf{KM Analysis of Progression-free Probability on TCGA Dataset\cite{cancer2013cancer}.} Group comparison by gender (right panel) is more robust than comparison by cancer type due to sample size per group (middle panel). Besides, the model interpretation of such a non-parametric is low since it only provides an overall progression-free chance on the entire cohort (left panel).}
    \label{fig:KM_estimation}
\end{figure}

I realize that KM-based survival models have three major limitations, which are given as follows:
\begin{enumerate}
    \item \textbf{Censoring and proportional hazards assumption}: One of the major limitations of KM analysis is the Censoring of data. Censoring occurs when the event of interest has not occurred in some participants or their follow-up time ends before the event occurs. This leads to incomplete information and may affect the accuracy of the results. Besides, KM analysis assumes that the hazard ratio is constant over time, which may not always be true. If the hazard ratio changes over time, the results of the KM analysis may not be reliable.
    \item \textbf{Sample size and group comparison}: The sample size can affect the accuracy and reliability of the results. If the sample size is too small, the results may not be statistically significant. Moreover, KM analysis can compare survival curves between different groups. However, the comparison may not be appropriate if the groups have different baseline characteristics that may affect the event of interest (\textbf{Figure}~\ref{fig:KM_estimation}, middle and right panel).
    \item \textbf{Low level of model explainability:} The model cannot explain model inference via parameters since it is a non-parametric model (\textbf{Figure}~\ref{fig:KM_estimation}, right panel).
\end{enumerate}

\color{black}
\subsection{Learning The Non-Linear Dynamics in Tumor Biology}\label{section:non_linear_tumor}
Modeling the dynamic of a physical body can be done in two ways. The first approach uses classical mechanics, for which we can find any other dynamical quantity by finding the position function $x(t)$. The second way is using the quantum mechanics approach, for which we aim to find the quantum wavefunction $\psi_n(x,t)$\cite{griffiths2018introduction}. To generalize, the wavefunction of a particle can be written as $\Psi(x,t)$, which is given by solving the Schrodinger equation\cite{griffiths2018introduction}
\begin{equation}
    i\hbar \frac{\partial \Psi(x,t)}{\partial t} = -\frac{\hbar^2}{2m}\frac{\partial^2  \Psi(x,t)}{\partial x^2}+V(x,t)\Psi(x,t),
\end{equation}
where $V(x,t)$ is the potential energy function. In most cases, we assume that $V$ is independent of $t$, which yields
\begin{equation}
    \Psi(x,t) = \psi(x)e^{-iEt/\hbar},
\end{equation}
where $E$ is the total energy, and $\hbar$ is Planck's constant.
\color{black}

\begin{figure}[t]
    \centering
    \includegraphics[width =0.8\textwidth]{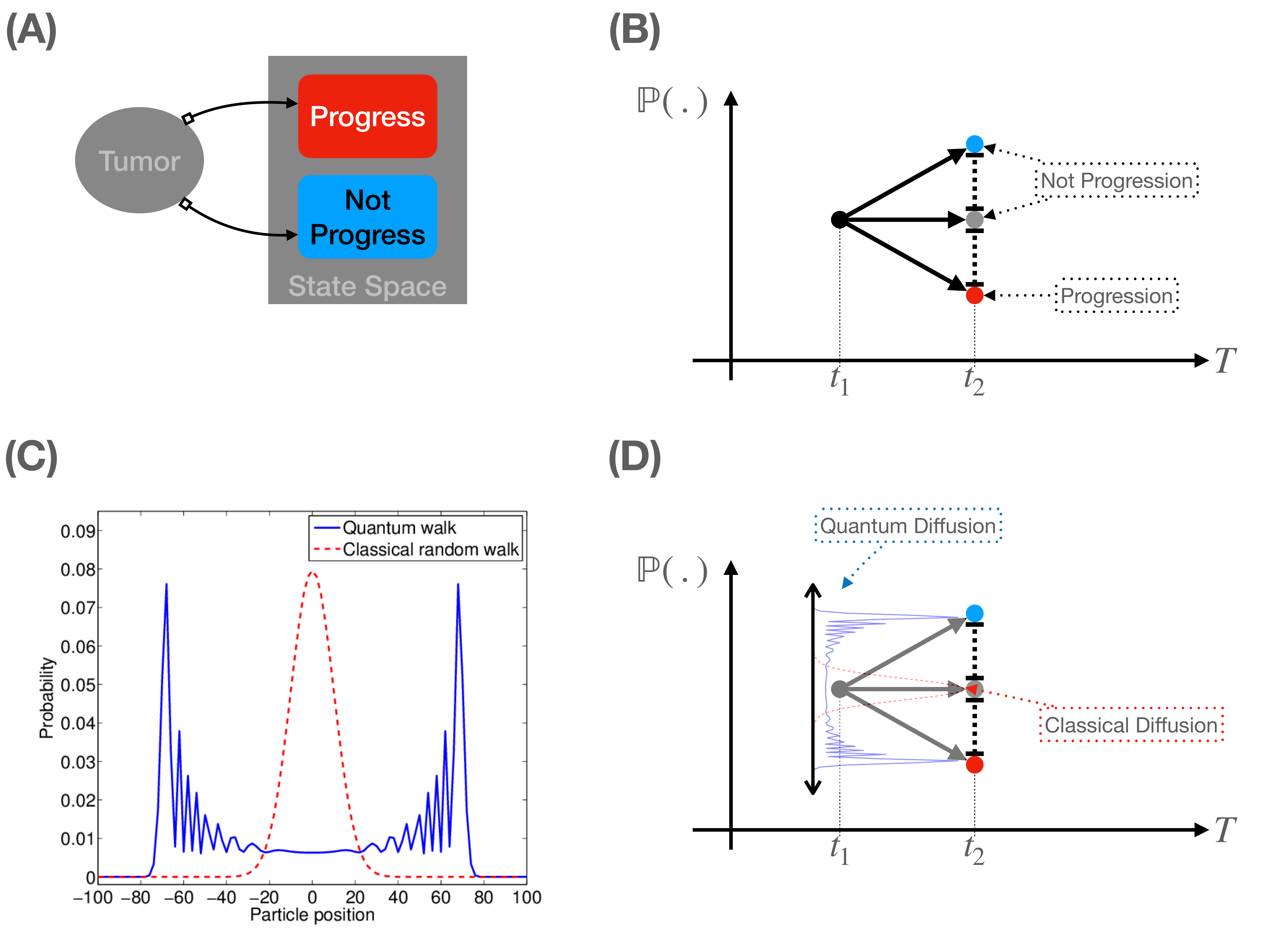}
    \caption{\textbf{Tumor Progression as Classical and Quantum Random Walks.} \textbf{(A)} Tumor progression state space, in which each tumor has only two future possibilities: (1) progression and (2) not progression. \textbf{(B)} Prediction surface of progression-free probability associated with tumor state space. \textbf{(C)} Classical v.s. Quantum Walks, figure credit from\cite{chandrashekar2010discrete}. \textbf{(D)} Progression-free probability interpreted by quantum and classical diffusion.}
    \label{fig:cls_quan}
\end{figure}

\color{black}
We find that some complex dynamics, such as tumor evolution, could be expressed in the term of quantum objects, which is depicted in \textbf{Figure}~\ref{fig:cls_quan}.
\begin{enumerate}
    \item Given a tumor $T$, treated as a physical system. The state-space of such a tumor is either progressing with the probability $p$ or not progressing as $(1-p)$ (\textbf{Figure}~\ref{fig:cls_quan}(A)).
    \item The prediction surface of progression-free probability associated with the state space is given in \textbf{Figure}~\ref{fig:cls_quan}(B). Specifically, progression-free survival (PFS) refers to \textbf{the period, including during and after the treatment of a disease, during which a patient lives with the disease, but it does not worsen}. The measurement of PFS probability (in [0,1]) in a clinical trial is used to determine the efficacy of a new treatment. Our interpretations of this implication are:
    \begin{enumerate}
        \item The progression-free probability of a disease is directly related to tumor progression. In cases where the progression-free probability is high (\textbf{Figure}~\ref{fig:cls_quan}(B)), it could imply that the disease is not advancing.
        \item Thus, the current treatment effectively controls or slows its growth. On the other hand, when the progression-free probability is low, it could imply that the tumor is growing and advancing despite the current treatment. 
    \end{enumerate}
    \item The change in PFS probability with stochastic processes, which can be classical or quantum random walks (\textbf{Figure}~\ref{fig:cls_quan}(C) and (D)).
\end{enumerate}

\color{black}
\subsection{A Particle Walking on 1D Lattice with An Approximate Trap}\label{sec:preliminary}
Previous work\cite{krapivsky2014survival} proposed a theoretical model to explain the quantum behavior of a quantum particle walking on a 1D lattice with the presence of a trap: Given a particle propagating on a 1D lattice with an initial position of $\alpha$ and a trap placed at the origin. The strength of the trap is tunable with parameter $\gamma$; i.e., we can amplify the trap's strength to entice the particle. With the trap, the wavefunction $\ket{\psi_{[.]}(t)}$ of the quantum particle located at site $n$ at time $t$ follows the time-dependent reduced tight-binding equation:
\begin{equation}\label{equa:survival}
    i\frac{d\ket{\psi_n(t)}}{dt} = \ket{\psi_{n+1}(t)} + \ket{\psi_{n-1}(t)} -i\gamma \delta_{n0}\ket{\psi_0(t)},
\end{equation}
where $\delta_{0n}$ is Dirac delta function. 

\begin{figure}[t]
    \centering
    \includegraphics[width = 0.8\textwidth]{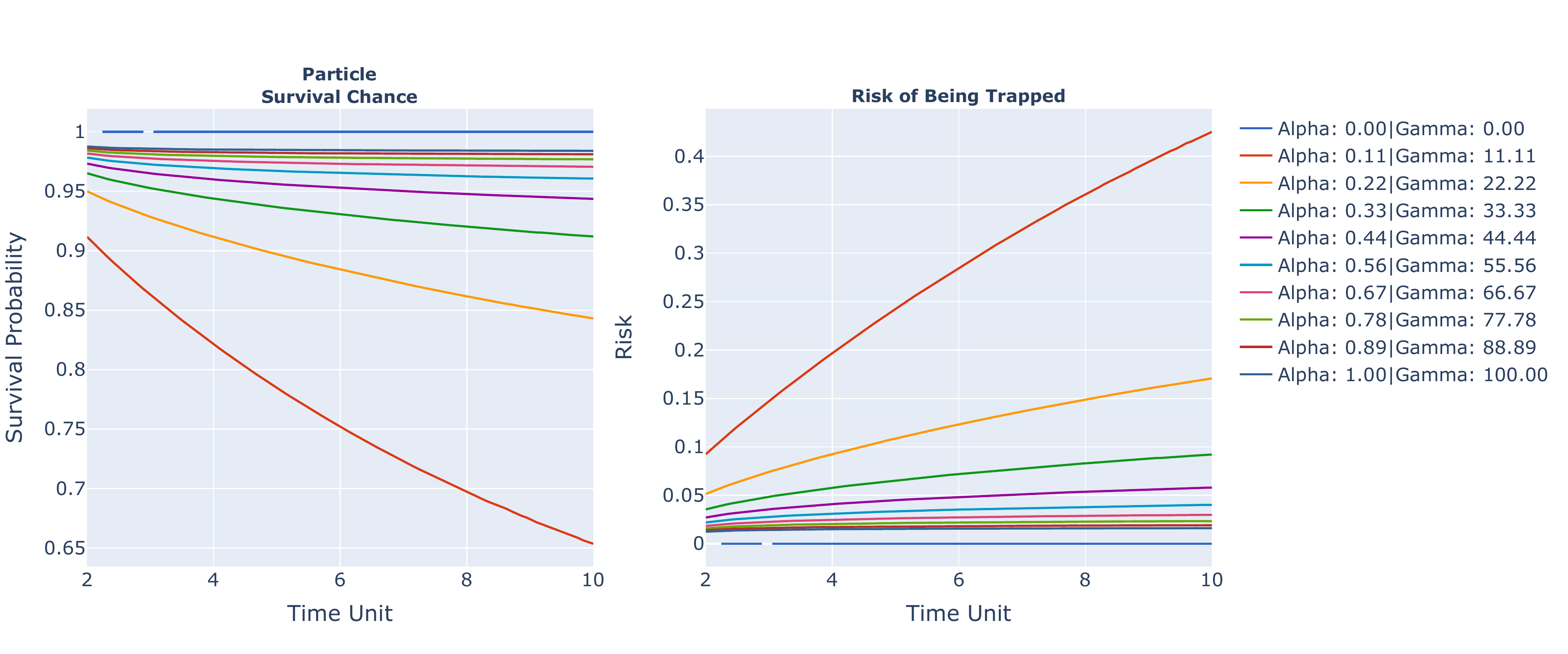}
    \includegraphics[width = 0.8\textwidth]{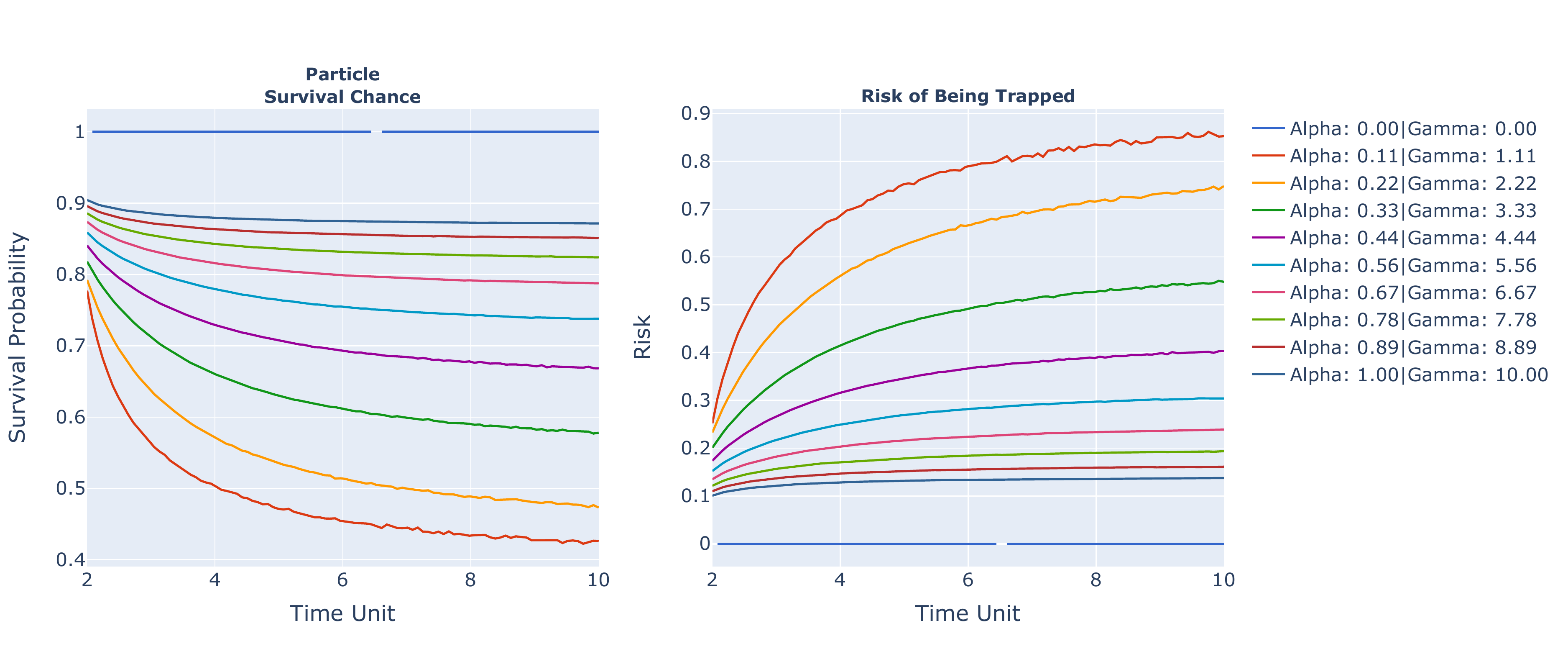}
    \includegraphics[width = 0.8\textwidth]{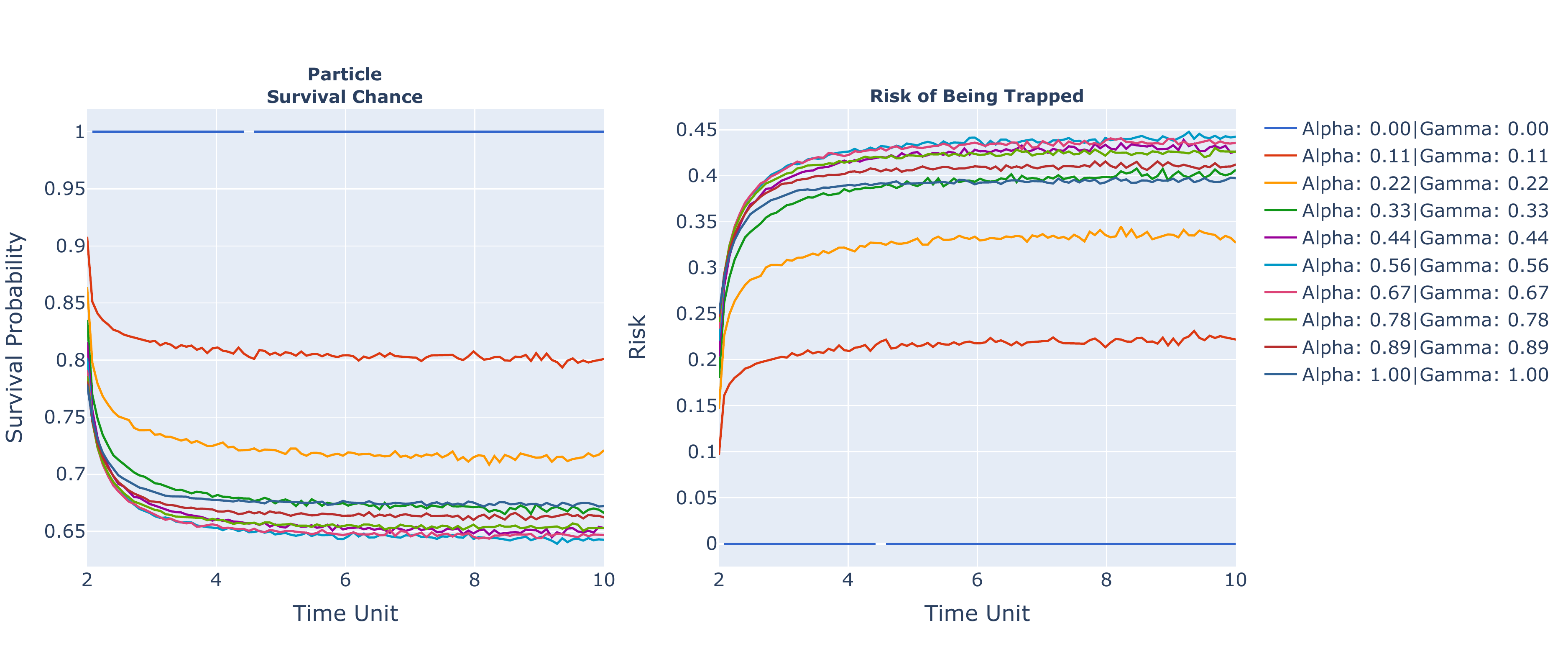}
    \caption{\textbf{Survival Probability of Quantum Particle Walking on 1D Lattice with the Presence of Proximate Trap.} I show the decreasing trap strength from the range [0,100] (top), [0,10] (middle), and [0,1] (bottom).}
    \label{fig:particle_trap}
\end{figure}

The physical phenomenon induces a paradoxical observation: the survival probability of a particle can increase as the strength of the trap increases, even if the particle is proximate to the trap. Specifically, the top panel of \textbf{Figure}~\ref{fig:particle_trap} shows the dynamics with extremely large trap strength of $\gamma \in [0,100]$ placed close to the quantum particle with $\alpha \in [0,1]$. It is observed that with increasing trap strength, the particle has an increased chance to be survived, which is paradoxical (Quantum Zeno effect\cite{krapivsky2014survival}) since the particle should have a greater chance of being trapped (not survive) in this scenario.
\color{black}

\section{Methods}\label{sec:cls_quantum}
\subsection{Treatment-Response Modeling by Enticing Quantum Particle with Proximate Trap}\label{sec:model}
We associate the initial location $\alpha$ of quantum particles with \textbf{patient indicators} (normalized in $[0,1]$) and the trap strength $\gamma$ as the \textbf{treatment effect}, with the model discussed in Section~\ref{sec:preliminary}. 

Of note, the discussed paradoxical phenomenon makes sense in the treatment-response model:
\begin{enumerate}
    \item Increasing the trap strength means amplifying treatment effects and increasing particle survival chances.
    \item By associating the particle survival with PFS, it can be interpreted that increasing treatment effect leads to higher PFS, and so the tumor does not progress (\textbf{Figure}~\ref{fig:cls_quan}(B)). 
\end{enumerate}

\textbf{Survival Probability and Risk Functions:} The time-dependent survival probability function of the particle or tumor concerning trap strength $\gamma$ is given by\cite{krapivsky2014survival}
\begin{equation}
    S(t|\gamma, \alpha) = 1 - 2\gamma \int_{0}^{t} |\ket{\psi_{0}(u)}|^2 du = 1-2\gamma I = A(\gamma) + B(t|\gamma,\alpha),
\end{equation}
where $|\ket{\psi_{0}(u)}|^2 = \langle\psi_{0}(u)|\psi_{0}(u) \rangle$ is the statistical interpretation of particle wavefunction\cite{griffiths2018introduction}. The main result of the work\cite{krapivsky2014survival} shows that such probability can be computed by using 
\begin{equation}
    I = A(\gamma) + B(t|\gamma,\alpha)
\end{equation}
where
\begin{equation}
    A(\gamma) = \frac{1}{\pi} \int_{0}^{2}\frac{du}{(\gamma + \sqrt{4-u^2})^2}
\end{equation}
and
\begin{equation}
    B(t|\gamma, \alpha) = \frac{1}{\pi} \int_{2}^{t} \frac{du}{\gamma^2 + u^2 -4} \bigg( \frac{u-\sqrt{u^2 - 4}}{2} \bigg)^{2\alpha}
\end{equation}

\subsection{Loss Module}\label{sec:loss_module}
We assume that for a quantified cohort of interest, there is an optimal value for the two parameters $(\gamma,\alpha)$ that yields the maximum likelihood for this data of interest. Specifically, we formulate for the loss function as
\begin{equation}\label{equa:CI}
    \mathcal{L}(\gamma,\alpha) = \text{ConcordanceIndex}(\rho(t), I_i)
\end{equation}
with input targets $I_i=(t_i,e_i)$, $t_i$ is the monitored time and $e_i$ is the indicator of progression/not-progression.

We propose a metric space to measure the treatment response score of an individual patient, given as
\begin{enumerate}
    \item Conventional survival analysis concerns the survival time of a patient, which coins the terms risk or hazard rate of this patient.
    \item In contrast, my work examines the tumor's survival concerning certain treatment effects
    \item As a result, the term "risk" becomes the tumor risk under the given treatment, which can be hypothesized to correlate negatively with patient risks.
    \item This biologically makes sense since the risk of the tumor getting cured (trapped) should be highly associated with the response effect of treatment on the patients.
    \item Hence, we denote the approximated risk of a tumor, i.e., response scores of patients as 
        \begin{equation}
            \rho(t) = \lambda(t) = -\log S(t).
        \end{equation}
\end{enumerate}
It is worth noting that the proposed patient response score $\rho(t)$ will be positive real numbers as the negative logarithm of value in $[0,1]$. Besides, the proposed model requires time calibration $t_\text{model}=t_\text{actual}+2$ to avoid non-sense computation of term $B(t|\gamma,\alpha)$ in the survival function; i.e., the upper-bound of the integral must be greater than lower-bound (here, we assume the tumor dynamics is moved forward in the timeline).

\subsection{Implementation}\label{sec:implementation}
We employ objective-oriented surrogate optimization\cite{bergstra2011algorithms} to produce Bayesian classifiers for the quantified problem, similar to my previous work\cite{nguyen2022bayesian,nguyen2022quantum}. Specifically, I assume that for a given survival data set $D= \{I_i\}$, there will be optimal $(\gamma,\epsilon_\gamma)$ and $(\alpha,\epsilon_\alpha)$, such that with
\begin{equation}
\begin{split}
        \gamma \leftarrow \mathcal{U} [\gamma- \epsilon_\gamma, \gamma+ \epsilon_\gamma]\\
        \alpha \leftarrow \mathcal{U} [\alpha- \epsilon_\alpha, \alpha+ \epsilon_\alpha]
\end{split}
\end{equation}
yields maximum concordance index (c-index) in \textbf{Equation}~\ref{equa:CI} with $\mathcal{U}$ is the uniform distribution. The Bayesian optimization with Tree Parzen Estimator is used for model optimization, which enables efficient model hyper-parameters and parameter optimizations in massive search space\cite{bergstra2011algorithms}.

\textbf{Pseudo-Code: Our Implementation for The Quantum Survival Model}
\begin{lstlisting}[language = Python]
import torch
import torch.nn as nn
import numpy as np
from torchquad import Monte-Carlo, set_up_backend

# UNIFORM DIST
def surv_prob(t, gamma, eta, return_log = True, debug = False):
    set_up_backend("torch", data_type="float32")
    mc = Monte-Carlo()
    if debug:
        print('t:     {}'.format(t))
        print('gamma: {}'.format(gamma))
        print('eta :  {}'.format(eta))
    def A(t):
        x = 1/(gamma + (4-t**2)**0.5)**2
        return x.requires_grad_()
    def B(t):
        x = (1/(gamma**2 + t**2 - 4))*((1/2*(t-(t**2-4)**0.5))**(2*eta))
        return x.requires_grad_()
    def _integrate(f, a, b):
        x = mc.integrate(f, dim=1, N= int(1e5),integration_domain=[[a, b]],backend="torch").requires_grad_()
        return x
    A_int = _integrate(A,0,2)
    B_int = _integrate(B,2,t)
    S = 1 - 2/np.pi*gamma*(A_int + B_int)
    if return_log:
        logS = torch.log(S)
        return S, logS
    else:
        return S
\end{lstlisting}

\section{Numerical Validation}\label{sec:result}
\begin{figure}[t]
    \centering
    \includegraphics[width = \textwidth]{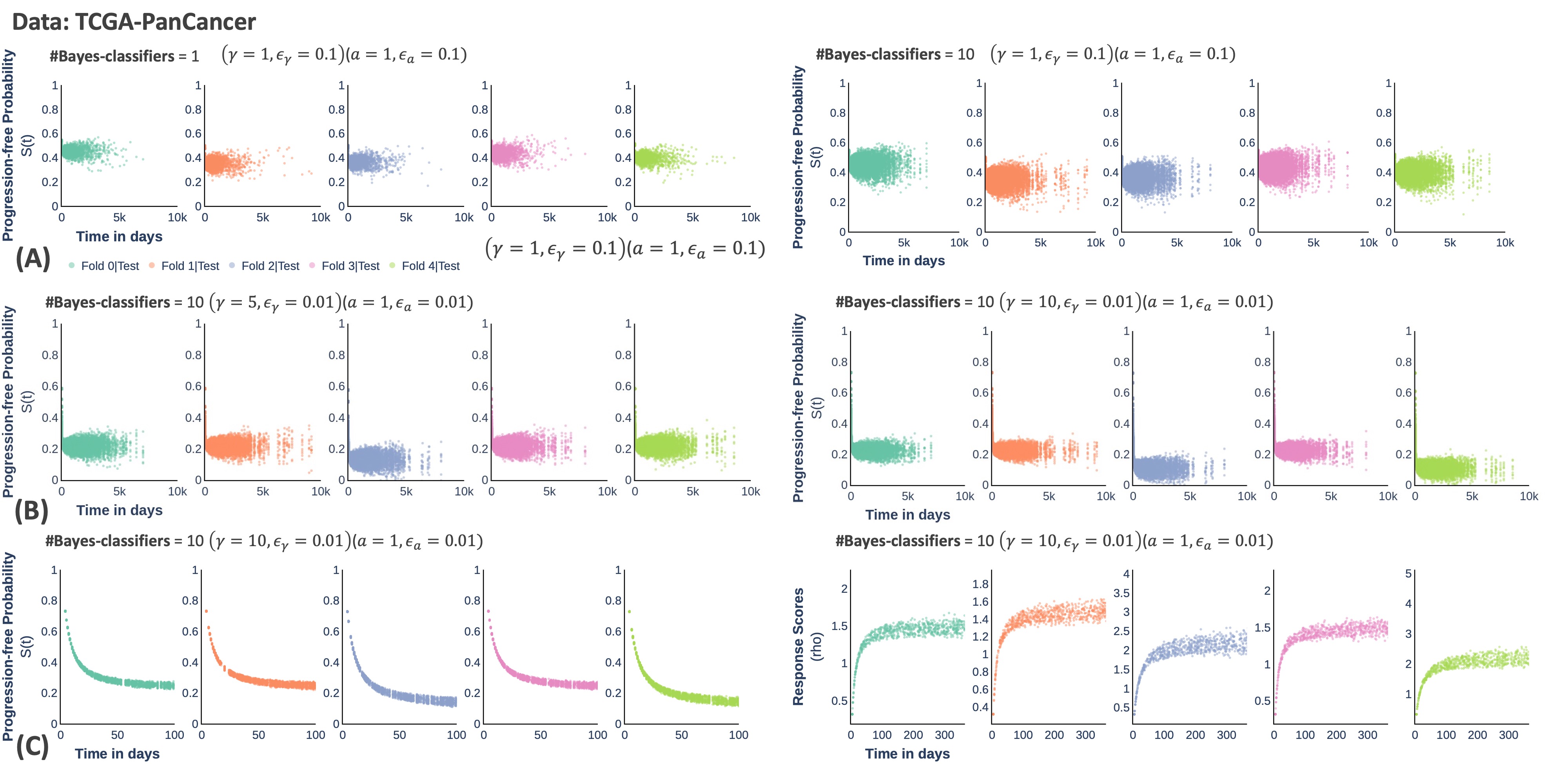}
    \caption{\textbf{The Bayesian Inference of Proposed Model for PFS probability of Pan-Cancer Analysis from TCGA Database.}}
    \label{fig:tcga_pancancer}
\end{figure}
\begin{figure}[t]
    \centering
    \includegraphics[width = \textwidth]{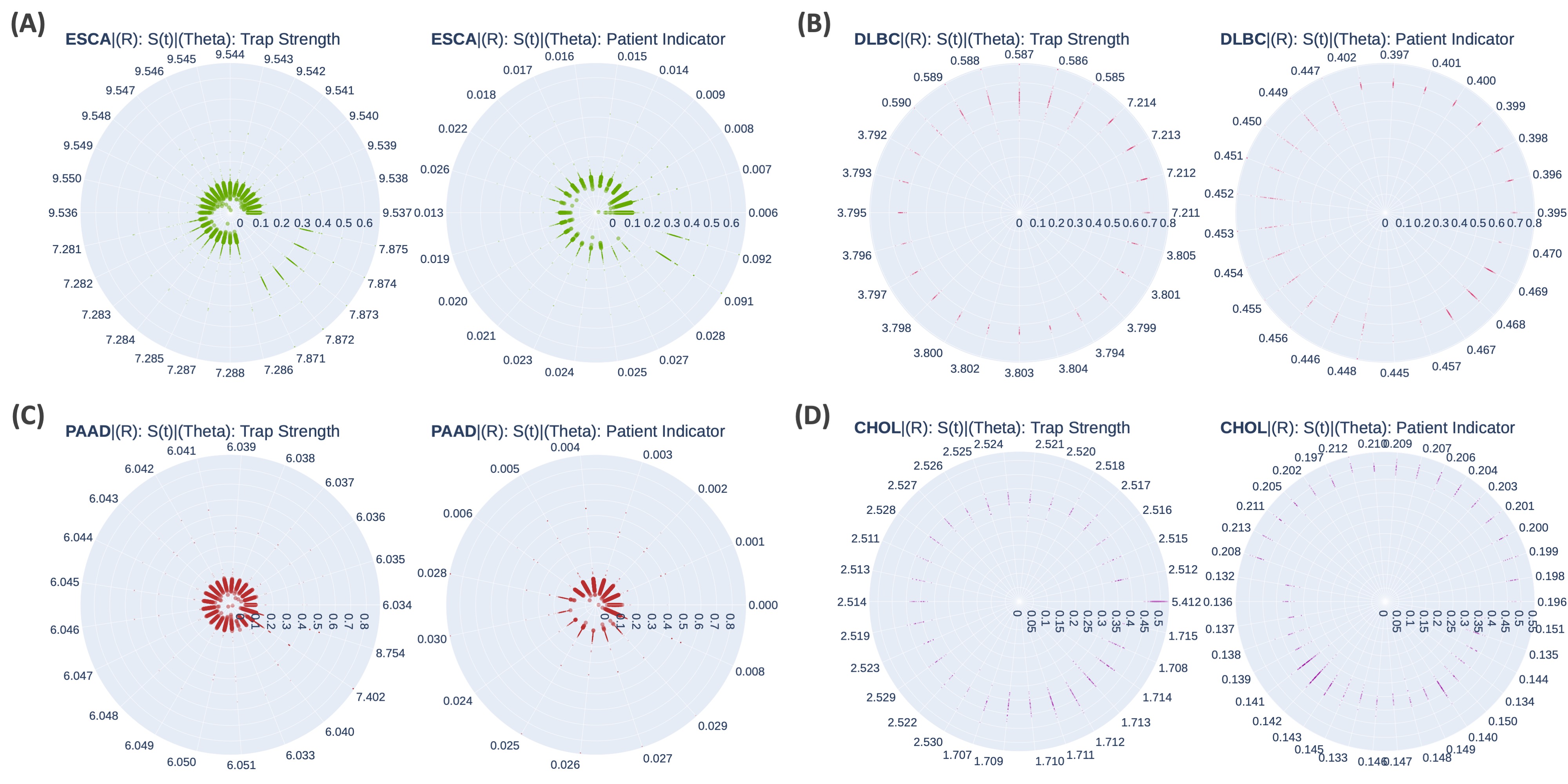}
    \caption{\textbf{Disease-Specific T.A.R.G.E.T Plots for Model Interpretation.}}
    \label{fig:target_plot_disease}
\end{figure}
\begin{figure}[t]
    \centering
    \includegraphics[width = 0.45\textwidth]{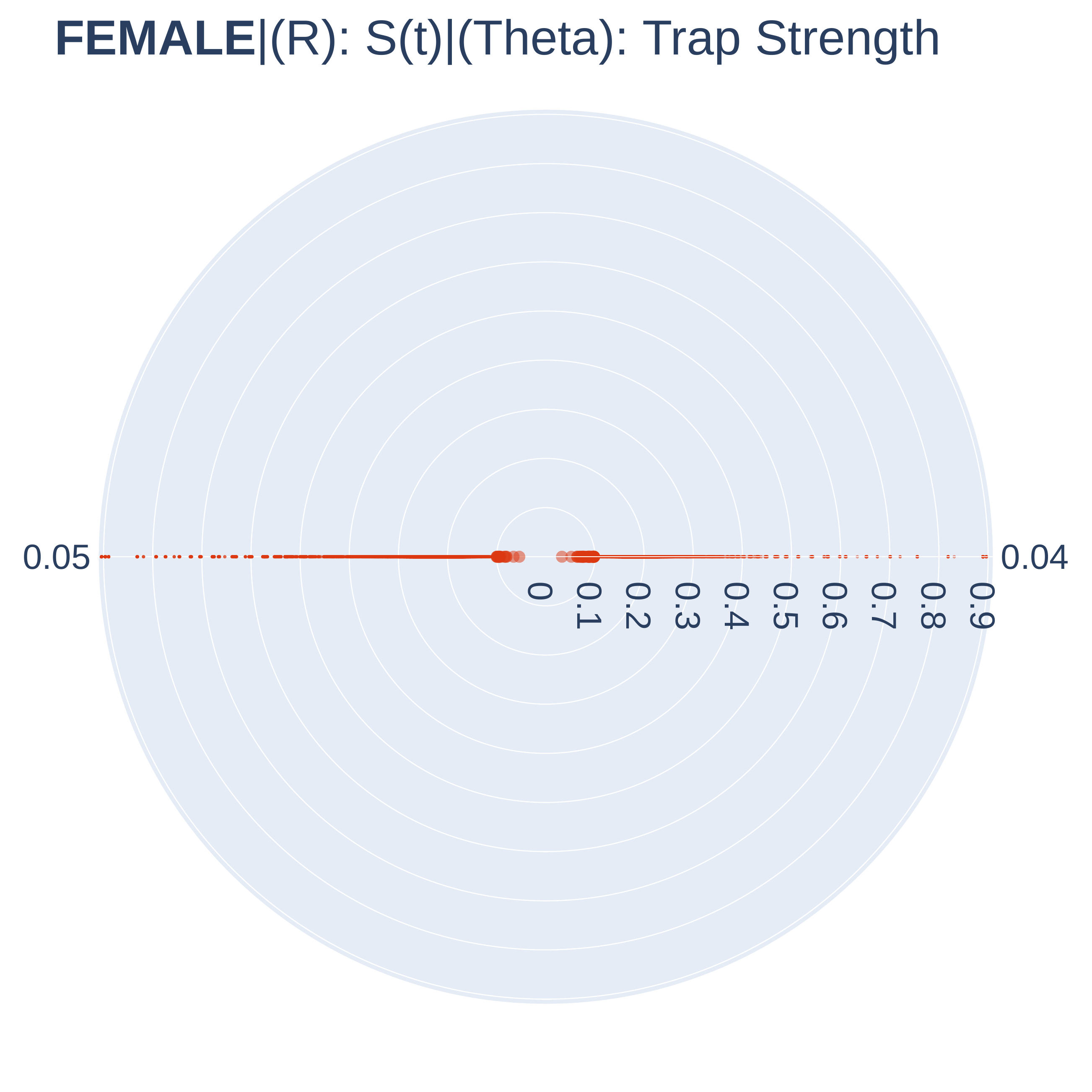}
    \includegraphics[width = 0.45\textwidth]{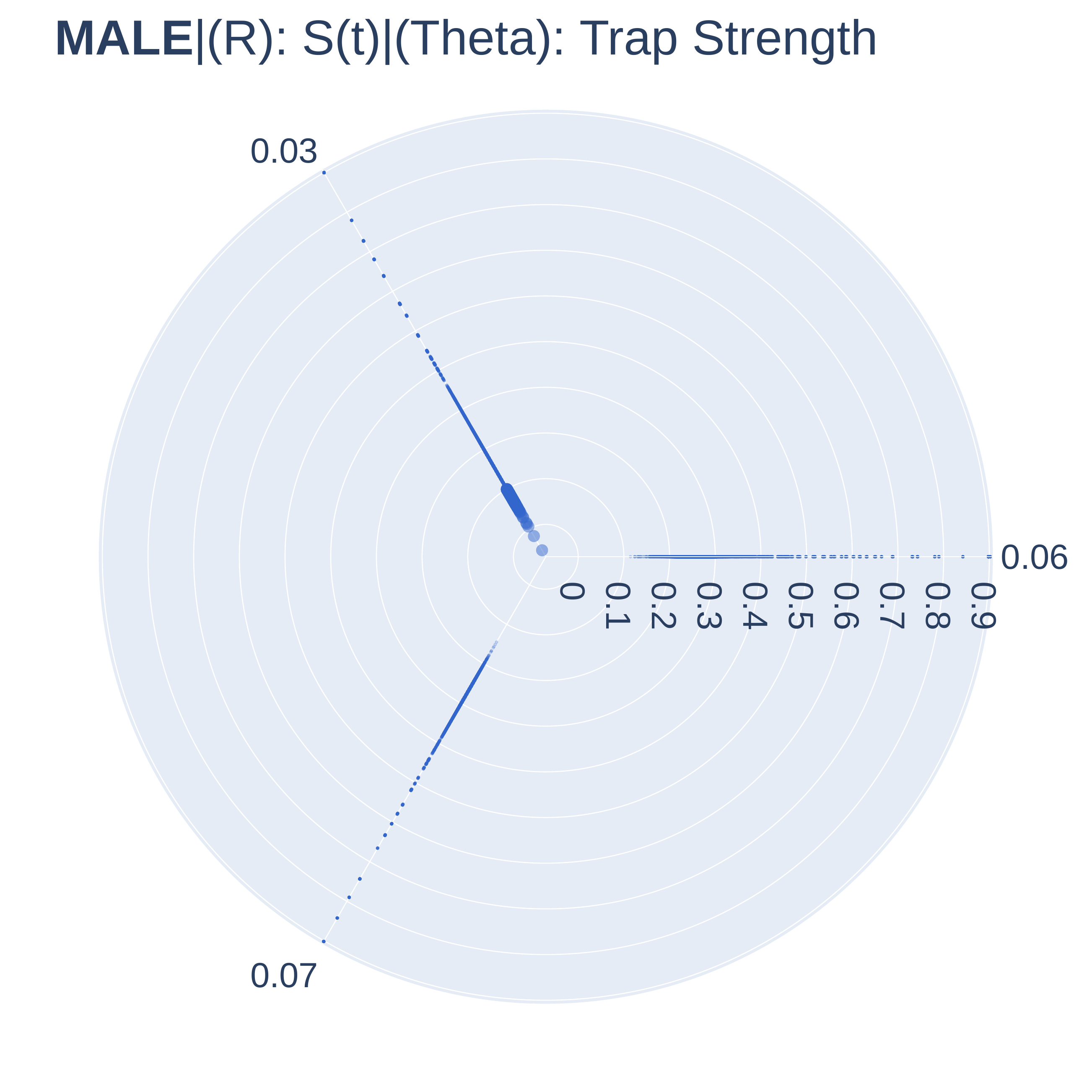}
    \caption{\textbf{Gender-Specific T.A.R.G.E.T Plots for Model Interpretation.}}
    \label{fig:target_plot_gender}
\end{figure}

\subsection{Environmental Setting}
Implementing the proposed quantum algorithm efficiently enables training using Graphical Processing Units (GPU). Besides, the experimental environment is limited computational hardware, with GeForce GTX $1060$ Mobile GPU and 6GB memory. All experiments was carried out using Python $3.7.0$, numpy $1.21.5$, scikit-learn $1.0.2$, and PyTorch $1.11$,
\color{black}

\subsection{Case-study}
The numerical result is demonstrated on The Cancer Genome Atlas databases\cite{cancer2013cancer}, with $10,907$ fully reported PFS data. The event indicator is $1$ for a patient having a new tumor, which can be due to disease progressions, local recurrence, distance metastasis, or a new primary tumor appearing in any site; otherwise, the data is censored. Of note, the data provider recommends the PFS for a relatively short follow-up time. Thus, we will use \textbf{days} for the unit of time in further analysis. Besides, I validate the model by 5-fold cross-validation, meaning the model is trained on $80\%$ of the data and tested on the separated remaining $20\%$ of the given set.

\subsubsection{Pan-Cancer Analysis of Progression-free Probability}
We report the Bayesian inference of the proposed model for PFS probability of Pan-Cancer analysis using the TCGA database in \textbf{Figure}~\ref{fig:tcga_pancancer}. We produce an arbitrary number of classifiers derived from the learned distribution of model weights to amplify the confidence of model prediction as $1$ and $10$ classifiers. With the same set of model parameters, more classifiers $(n=10)$ propose larger confident regions for model prediction, which covers the case of smaller number classifier $(n=1)$. Besides, the distribution is shifted, corresponding to increasing trap strength from $5$ to $10$ (conceptualized) units, depicted in the middle panel.

The quantum diffusion shows a strangeness in the proposed model prediction surface. Specifically, the proposed model prediction suggests the PFS probability of the patient is aggressively reduced within only $100$ days after the monitored time. After this time frame, the model suggests that the predicted chance of tumor progression is randomly distributed surrounding certain values, ranging from $0.2-0.4$.

Although showing the stranger dynamics due to quantum diffusion, the model prediction is consistent with the KM model (\textbf{Figure}~\ref{fig:KM_estimation}). Specifically, after $5,000$ days, the model in \textbf{Figure}~\ref{fig:tcga_pancancer}, top-left panel suggests the PFS chance is around 50\%, slightly higher than the KM model, which means such probability beyond $5,000$ days is under roughly 40\% (\textbf{Figure}~\ref{fig:KM_estimation}). The gap between the two models is closed as the monitored time reaches the endpoint. Specifically, the KM model suggests that such probability is around $0$ to $0.2$ at approximately $10,000$ days ($\approx 27$ years). The same observations can be drawn from the middle panel of \textbf{Figure}~\ref{fig:tcga_pancancer}, in which model predictions at the endpoint are randomly diffused around $0$ in the second fold with trap five units; $0.2$ in the second and the myth fold with trap ten units; and $0.2$ in the remaining folds.

By linking the PFS chance with response scores via negative correlations, we can estimate the response scores of the entire TCGA PanCancer cohort forward in time. Specifically, the proposed model suggests that the response effect of treatment may significantly increase from $0$ to approximately $[1.2-2.0]$. At the diffusion time after $100$ days, the response effects are saturated and plateau at some values. These plateau effects biologically make sense since treatment effects could decay as the drug is decayed by exposure to the tumor. For example, the treatment response score stops growing significantly after $100$ days of the first year, and this effect gradually increases within the remaining $256$ days. Using relatively short time-to-event data can be a pattern to construct treatment-response modeling for long-term treatment planning. Specifically, given two observations collected on days $100$ and $200$, I can estimate the response score by the end of the day $365$ by knowing the functional surface's general dynamics representing model predictions. 

\subsubsection{Sub-Group-Specific Analysis}
\textbf{Disease-specific Analysis:} We propose T.A.R.G.E.T plots to interpret the proposed quantum survival model in \textbf{Figure}~\ref{fig:target_plot_disease} and ~\ref{fig:target_plot_gender}. We hypothesize that the tumor (particle) survival chance can be modeled via the quantum mechanism of trapping quantum particles on the 1D chain. We link this physically-inspired mechanism to the plot,
\begin{itemize}
    \item A trap is placed at the center of the plot, enticing particles coming from different configurations of models.
    \item This variational learning explains model prediction via the two proposed-hypothesized indicators, including trap strength (presented on the radial unit (theta) surrounding the plot).
    \item The proposed model interpretation reveals intrinsic patterns from the survival chance of tumors stratified by diseases. Samples with PFS chance below $0.1$ is depicted in bigger markers (\textbf{Figure}~\ref{fig:target_plot_disease}). 
    \item ome cancer phenotypes present better response scores to treatment, i.e., higher PFS, like DLCB or CHOL. In contrast, ESCA and PAAD have more aggressive behaviors to treatment since more samples with PFS probability below $0.1$.
\end{itemize}

\textbf{Gender-specific Analysis:} We can use a T.A.R.G.E.T plot to compare the behavior of tumors stratified by genders in \textbf{Figure}~\ref{fig:target_plot_gender}. The proposed model suggests that female tends to have a narrow range for the trap strength to reduce the PFS below $0.1$ effectively, ranging from $0.04$ to $0.05$ with discretized two digits after the fractions. In contrast, male patients are stratified into two subgroups, effectively treated at $0.03$ and ineffective regions between $0.06$ and $0.07$. This observation suggests that treatments for females require stronger effects (larger trap strength) than male treatment to reduce the PFS chance of tumor below $0.1$. 
\color{blue}
We emphasize that this is the implication derived from observing the prediction surface of the proposed T.A.R.G.E.T plots and should not be applied in any clinical context.
\color{black}

\color{black}
\section{Discussion and Conclusion}\label{sec:discussion}
\subsection{Reproducible Capacity of The Works}
The train-test split of our numerical result is given in the GitHub repository: "\url{https://github.com/namnguyen0510/Tumor-As-Quantum-Particle/tree/main}". We deliberately save the data for each fold to reproduce the work. The experimental history is saved in the database file ".db" for model inference. The train and test fold are completely separated.

\subsection{Ethical Implications}
We strongly suggest not to apply this model in any clinical context. The main goal of this work is to test whether the quantum model is relevant to biological applications. Most of the current approaches\cite{kuang2018introduction} take the classical perspectives of a physical system. In this work, the underlying assumption is that tumor dynamics could behave as a quantum object, for which its PFS could be explained in quantum-statistical mechanics.

\subsection{Conclusion}
To this end, we have presented a quantum survival model in the context of treatment response modeling and outlined the theory and concept behind it in \textbf{Sections}~\ref{sec:cls_quantum} and ~\ref{sec:model}. An efficient training protocol is introduced on advanced computing hardware in \textbf{Section}~\ref{sec:implementation}. We have applied the model to a practical case study involving Big Data of Pan-Cancer PFS analysis to demonstrate the proof of concept, as presented in \textbf{Section}~\ref{sec:result}.

\section{Acknowledgement}
The author would like to thank all colleagues for stimulating and constructive comments.

\bibliography{output}
\bibliographystyle{abbrv}

\begin{appendices}
    
\end{appendices}

\lstset{style=mystyle}

\end{document}